\documentstyle[aps,epsf,prl,multicol]{revtex}

\begin{document}
\title{Decoherence of Atomic Gases in Largely Detuned Laser Fields}

\author{Karl-Peter Marzlin }
\address{Fachbereich Physik der Universit\"at Konstanz, 
   Postfach 5560 M674, D-78457 Konstanz, Germany}

\date{\today}
\maketitle
\begin{abstract}
We study theoretically the decoherence of a gas of bosonic atoms
induced by the interaction with a largely detuned laser beam. It is
shown that for a standing laser beam decoherence coincides
with the single-particle result. For a running laser beam
many-particle effects lead to significant modifications.
\end{abstract}

$ $ \\
32.80.-t, 03.65.Yz, 03.75.Fi
\begin{multicols}{2}
In experiments with atomic Bose-Einstein
condensates \cite{parkins98,stringari99} the trap which is used to isolate
the atoms from the environment plays an important role for the 
physical behaviour of the system. In a magnetic trap 
usually only atoms with a specific magnetic hyperfine quantum number
are stored. An optical trap \cite{ketterle98} 
can be used to confine atoms in all
Zeeman sublevels. However,
in order to preserve the atomic coherence 
spontaneous emission of photons must be avoided.
A large detuning $\Delta$ of the laser beams providing the trap field
reduces the excitation probability for atoms and thus reduces
the number of spontaneously emitted photons. At the same time,
a relatively strong intensity of the laser beams, and correspondingly
a large Rabi frequency $\Omega ({\bf x})$, ensures that the laser
beams' effect is still large enough to produce a strong potential
for the atoms. 

The aim of this paper is to examine light-induced decoherence of
a gas of ultracold atoms in an optical trap and to deduce
the influence of atomic many-particle properties on this
decoherence. To do so we study solutions to a master equation
for reduced Heisenberg operators describing a quantum
field of bosonic two-level atoms interacting with a detuned laser beam
and the vacuum fluctuations of the electromagnetic field.
The technique of reduced Heisenberg operators has previously been
employed by Nemoto and Shibata \cite{nemoto96} and allows an
elegant description of decoherence in a many-particle context.
After the derivation of the master equation we will adiabatically
eliminate the excited state to get an effective master equation for
the atomic internal ground state. 
The main result of the paper is that many-particle effects lead
to a nonlocal modification of light-induced decoherence that
vanishes for laser fields with spatially homogeneous phase
(e.g., a standing laser wave) but makes substantial contributions
for a running laser wave.

{\bf Derivation of the master equation:}
We consider a general system that is described by the direct product
of two Hilbert spaces ${\cal H}_S$ and ${\cal H}_R$, where
${\cal H}_S$ describes the open system that we are interested in, and
${\cal H}_R$ describes the reservoir to which the system is coupled. 
It is our aim to find a master equation describing the evolution of reduced
Heisenberg operators for the system only. We define a
reduced Heisenberg operator $\hat{R}_S$ by the relation
\begin{equation} 
   \hat{R}_S(t) := \mbox{Tr}_R\{\hat{\rho}_R \hat{R}(t)\} \; ,
\end{equation} 
where $\hat{R}(t)$ is the original Heisenberg operator which acts
on  ${\cal H}_S \otimes {\cal H}_R$ and $\hat{\rho}_R$ is the
density matrix describing the state of the
reservoir. It is thereby assumed that the (time-independent) 
density matrix can be written in the form $\hat{\rho} =
\hat{\rho}_S \otimes \hat{\rho}_R$ so that there are initially no
correlations between the system and the reservoir. The physical
significance of the reduced Heisenberg operator is that it allows
to describe accurately any measurement involving the system only,
i.e.,
\begin{equation} 
   \mbox{Tr}_{S\otimes R}
   \{ \hat{\rho} \hat{F}_S \hat{R}(t) \hat{F}_S^\prime \} =
   \mbox{Tr}_S\{\hat{\rho}_S \hat{F}_S \hat{R}_S(t) \hat{F}_S^\prime \}
\end{equation} 
for any two operators $ \hat{F}_S$ and  $ \hat{F}_S^\prime $ acting
on ${\cal H}_S$ only.

To derive a master equation for $\hat{R}_S(t)$ we employ Zwanzig's method
by using that
${\cal P}\hat{R} := \mbox{Tr}_R\{ \hat{\rho}_R \hat{R} \} \otimes
{\bf 1}_R$ is a projection superoperator fulfilling ${\cal P}^2 ={\cal P}$.
We start from the Heisenberg equation-of-motion
\begin{equation} 
   i\hbar \partial_t \hat{R}(t) = [\hat{R}(t), \hat{H} ] 
   =: {\cal L} \hat{R}(t) \; ,
\end{equation} 
where ${\cal L}$ is the Liouville superoperator. 
The Hamiltonian in Heisenberg picture is explicitely
time-independent if it is written in terms of Schr\"odinger picture
operators. 
Introducing ${\cal Q} := 1 - {\cal P}$ and following
the lines of Ref.~\cite{mandlwolf} one can derive the
master equation for the reduced Heisenberg operator,
\begin{eqnarray}  
   i \hbar \partial_t \hat{R}_S(t) &=& {\cal P}{\cal L} {\cal P}
    \hat{R}_S(t) \label{master1} \\
   & &  -\frac{i}{\hbar} {\cal P} {\cal L} 
    \int_{t_0}^t dt^\prime U_{QQ}(t^\prime) {\cal Q}
    {\cal L} {\cal P} \hat{R}_S(t-t^\prime) \; .
\nonumber \end{eqnarray} 
In the derivation it was assumed that the inital Heisenberg operator
$\hat{R}(0)$ acts on the system only so that ${\cal Q}\hat{R}(0)=0$.
The superoperator
 $  U_{QQ}(t^\prime ) :=  \exp \{ -i t^\prime
       {\cal Q} {\cal L} {\cal Q}/\hbar \}$
describes the unitary evolution in ${\cal Q}$-subspace.

This expression can be further simplified if the Liouvillean is
written in the form ${\cal L} = {\cal L}_S + {\cal L}_R + 
{\cal L}_{\mbox{{\scriptsize int}}}$, where ${\cal L}_S$ and
${\cal L}_R$ act solely on the system and the reservoir, respectively,
and ${\cal L}_{\mbox{{\scriptsize int}}}$ describes the interaction.
It is then easy to show that ${\cal Q} {\cal L} {\cal P} = {\cal Q}
{\cal  L}_{\mbox{{\scriptsize int}}} {\cal P}$ and, if the reservoir
is in a stationary state ($[\hat{H}_R, \hat{\rho}_R]=0$), 
 ${\cal P} {\cal L} {\cal Q} = {\cal P}
{\cal  L}_{\mbox{{\scriptsize int}}} {\cal Q}$. Expanding 
Eq.~(\ref{master1}) to second
order in ${\cal L}_{\mbox{{\scriptsize int}}}$ we can replace
$U_{QQ}(t^\prime)$ by $\exp\{ -it^\prime ({\cal L}_S + {\cal
L}_R)/\hbar \}$. Performing a Markov approximation
($\hat{R}_S(t-t^\prime) \approx U_{PP }^{-1}(t^\prime) \hat{R}_S(t)$
under the integral)
then results in the final form of the master Equation in Heisenberg
picture,
\begin{equation}  
   i \hbar \partial_t \hat{R}_S(t) \approx {\cal L}_S \hat{R}_S(t)  
    -\frac{i}{\hbar} {\cal P} {\cal L}_{\mbox{{\scriptsize int}}}    
    {\cal Q} \int_{0}^\infty dt^\prime 
    {\cal L}_{\mbox{{\scriptsize int}}}(t^\prime )
    \hat{R}_S(t) \; ,
\label{mastergl} \end{equation} 
with ${\cal L}_{\mbox{{\scriptsize int}}}(t^\prime ) \hat{R}_S =
  [  \hat{R}_S, \hat{H}_{\mbox{{\scriptsize int}}}(t^\prime )]$ and
$ \hat{H}_{\mbox{{\scriptsize int}}}(t^\prime ) := 
  \exp \{ -it^\prime ({\cal L}_S + {\cal L}_R)/\hbar \}
     \hat{H}_{\mbox{{\scriptsize int}}}$.

{\bf Decoherence of the atomic field operator:}
The formalism of the previous section will now be applied to the problem
of spontaneous emission in a system of bosonic two-level atoms. 
In this case the degrees-of-freedom of the electromagnetic field
play the role of a reservoir to which the atoms are coupled.  Since
it is known that usually the center-of-mass motion of the atoms
has only a small influence on the spontaneous emission rate
\cite{rzazewski92,marzlin00} we neglect it alltogether and use
the atomic Hamiltonian
\begin{equation} 
  \hat{H}_S = \int d^3x \hat{\Psi}_e^\dagger ({\bf x}) \hbar \omega_0
         \hat{\Psi}_e ({\bf x}) \; ,
\end{equation} 
where $\omega_0$ is the resonance frequency and $ \hat{\Psi}_i ({\bf
x})$ are the quantum field operators for excited ($i=e$) and
ground-state ($i=g$) atoms, respectively. The Hamiltonian of the
reservoir is given by 
\begin{equation} 
  \hat{H}_R = \frac{\varepsilon_0}{2} \int d^3x \{ \hat{{\bf E}}^2 +
        c^2 \hat{{\bf B}}^2 \}  \; ,
\end{equation}
and the coupling between the system of two-level atoms and the
reservoir
is described by the electic-dipole coupling in rotating-wave
approximation,
\begin{eqnarray}  
  \hat{H}_{\mbox{{\scriptsize int}}}(t^\prime ) &=& \int d^3x \{ 
  \hat{{\bf E}}^{(+)}({\bf x},t^\prime ) \hat{{\bf P}}^{(-)}({\bf x}) 
  e^{i\omega_0 t^\prime } +
  \nonumber \\ & & \hspace{1.1cm} 
  \hat{{\bf E}}^{(-)}({\bf x},t^\prime) \hat{{\bf P}}^{(+)}({\bf x}) 
  e^{-i\omega_0 t^\prime } \} \; ,
\end{eqnarray} 
with $\hat{{\bf E}}^{(+)}({\bf x},t^\prime )$ being the positive-frequency
part of the electric field in interaction picture and
$\hat{{\bf P}}^{(+)}({\bf x}) := {\bf d} \hat{\Psi}_e ({\bf x})
\hat{\Psi}_g^\dagger ({\bf x})$ is the positive-frequency part
of the polarization operator.

Under the assumption that the initial state of the electromagnetic
field is the vacuum $|0 \rangle $ we can insert the various
Hamiltonians into Eq.~(\ref{mastergl}) to derive
\begin{equation} 
   i \hbar \partial_t \hat{R}_S(t) = [ \hat{R}_S(t) , \hat{H}_S]
   -i \hbar {\cal L}_D \hat{R}_S(t) \; ,
\label{2levmaster1} \end{equation} 
with the decoherence Liouvillean
\begin{eqnarray} 
   {\cal L}_D \hat{R} &:=&
   \int d^3x d^3x^\prime 
   \Big \{ 
   [ \hat{R} , P^{(-)}_i({\bf x}) ] P^{(+)}_j({\bf x}^\prime) 
   T_{ij}({\bf x}, {\bf x}^\prime ) +
   \nonumber \\ & & \hspace{4mm}
    P^{(-)}_i({\bf x}) [ P^{(+)}_j({\bf x}^\prime ), \hat{R} ]  
   T_{ij}^*({\bf x},{\bf x}^\prime )\;  \Big \} .
\label{ldec} \end{eqnarray} 
The quantity $T_{ij}$ 
can be calculated by standard methods
(see, e.g., Refs.~\cite{zhang94,marzlin98}) and is given by
\begin{eqnarray} 
  T_{ij} ({\bf x}, {\bf x}^\prime ) 
     &:=& \int_0^\infty dt^\prime 
     \langle 0 | E^{(+)}_i({\bf x}, t^\prime ) 
                 E^{(-)}_i({\bf x}^\prime , 0) |0 \rangle 
  \label{tij} \\ &=&
     \int_0^\infty\frac{ d\omega \omega^3}{4\pi^2 \hbar
     \varepsilon_0 c^3} 
     \left ( \pi \delta(\omega -\omega_0) -i
     \frac{P}{\omega -\omega_0} \right ) \times
  \nonumber \\ & &
     \Big [\delta_{ij} \left ( j_0(\frac{\omega r}{c}) - \frac{ 
     j_1(\frac{ \omega r}{c})}{\frac{\omega r}{c}} \right ) + 
  \nonumber \\ & &
    \hat{{\bf r}}_i \hat{{\bf r}}_j
    \left ( \frac{3 j_1(\frac{ \omega r}{c})}{\frac{ \omega r}{c}}- 
    j_0(\frac{ \omega r}{c})  \right ) \Big ] \; .
\nonumber \end{eqnarray} 
In the last expression we have defined $r:= | {\bf x}-{\bf x}^\prime|$ 
and $\hat{{\bf r}} := ({\bf x}-{\bf x}^\prime )/r$.

It is not hard to see that the imaginary part of $T_{ij}$ leads only
to a (non-local) modification of the Hamiltonian $\hat{H}_S$ and
therefore does not lead to decoherence effects. Since it is the latter
which we are interested in we will omit the imaginary part. Doing so
and inserting the definition of the polarization operator into
Eq.~(\ref{ldec}) results in
\begin{eqnarray}  
   {\cal L}_D \hat{R}  &=& \frac{3 \gamma_0}{4}  
   \int d^3x d^3x^\prime 
   \Big \{ 
   [ \hat{R} ,\Psi_e^\dagger({\bf x}) \Psi_g({\bf x}) ]
      \Psi_g^\dagger({\bf x}^\prime)  \Psi_e({\bf x}^\prime)  +
  \nonumber \\ & &
   \Psi_e^\dagger({\bf x}) \Psi_g({\bf x})  
      [\Psi_g^\dagger({\bf x}^\prime)\Psi_e({\bf x}^\prime),\hat{R}]
      \Big \} J(r, \theta)\; .
\label{2levmasterfin} \end{eqnarray} 
Here $\gamma_0 := |{\bf d}|^2 \omega_0^3 /(3\pi \hbar \varepsilon_0 c^3)$
is the single-atom spontaneous emission rate and
the function $ J(r, \theta)$ is defined as
\begin{equation}   J(r, \theta) := 
   \sin^2(\theta) j_0(\frac{\omega_0 r}{c}) +
   \frac{j_1(\frac{ \omega_0 r}{c})}{\frac{\omega_0 r}{c}}
   \left ( 3 \cos^2(\theta)-1 \right )\; ,
\end{equation} 
where $\theta$ is the angle between ${\bf d}$ and $\hat{{\bf r}}$. 
In the context of master equations for reduced density matrices the
decoherence term (\ref{2levmasterfin}) is well known 
\cite{lehmberg70,castin98}
and has recently been applied to study loading \cite{santos00a} and
condensation \cite{santos00b} of a condensate in laser light.

Eq.~(\ref{2levmasterfin}) will be the starting point of our
investigation on light-induced decoherence in many-atom systems.
Before doing so we point out that the master equations
Eq.~(\ref{2levmasterfin}) and (\ref{ldec}) are very similar to
corresponding equations for the reduced density matrix in
Schr\"odinger picture. However, there is a striking difference:
For the reduced density matrix $\hat{\rho}_S$ the Liouvillean for spontaneous
emission includes terms in which the operators
$\hat{{\bf P}}^{(+)}({\bf x}) $ and $\hat{{\bf P}}^{(-)}({\bf x}) $
appear to the left and to the right of the density matrix,
respectively. This ordering is important as such a term describes
the incoherent de-excitation of an atom, while the opposite
ordering $\hat{{\bf P}}^{(+)}({\bf x})  \hat{\rho}_S
\hat{{\bf P}}^{(-)}({\bf x}) $ would describe an incoherent excitation
of an atom. But it is exactly the latter type of ordering which
appears in Eq.~(\ref{ldec}). The reason is that Eq.~(\ref{ldec})
does not describe the evolution of the density matrix in Schr\"odinger
picture but that of an operator in the Heisenberg picture. When
applied to a Heisenberg operator the ``wrong'' ordering gives indeed
the correct time evolution.

{\bf A gas of atoms in largely detuned laser light:}
To study the decoherence of atoms in a largely detuned laser beam
we adiabatically eliminate the excited state by means of (see, e.g., 
Ref.~\cite{marzlin98b}) 
\begin{equation} 
  \hat{\Psi}_e({\bf x}) \approx \frac{\Omega({\bf x})}{\Delta} 
  \hat{\Psi}_g({\bf x}) \; .
\label{adiabelem}\end{equation} 
Inserting this into Eq.~(\ref{2levmasterfin}) leads to a Liouvillean
\begin{eqnarray}  
   {\cal L}_g \hat{R}  &=& \frac{3 \gamma_0}{4}   
   \int d^3x d^3x^\prime 
   \frac{\Omega^*({\bf x})\Omega({\bf x}^\prime)}{\Delta^2} 
   J(r,\theta) \times \nonumber \\ & &
   \Big \{ [\hat{R}, \hat{\rho}_g({\bf x})] \hat{\rho}_g({\bf x}^\prime )
   - \hat{\rho}_g({\bf x})[ \hat{R}, \hat{\rho}_g({\bf x}^\prime )]
   \Big \}
\label{ladiab} \end{eqnarray} 
with $ \hat{\rho}_g({\bf x}) := \hat{\Psi}_g^\dagger({\bf x})
\hat{\Psi}_g({\bf x})$.

Eq.~(\ref{ladiab}) describes the decoherence of any Heisenberg
operator acting on the ground state of the atoms in the presence
of a laser beam. Its physical interpretation is that the laser beam
excites an atom 
at position ${\bf x}^\prime$ to a virtual state. The atom
then spontaneously emits a photon which is absorbed by another atom
at position ${\bf x}$. Finally, this atom makes an induced emission
of a photon into the mode of the laser beam. 
A simple application of Eq.~(\ref{ladiab}) is the demonstration
of lack of decoherence
for the number density operator $ \hat{\rho}_g({\bf x})$. Since
$[\hat{\rho}_g({\bf x}) ,\hat{\rho}_g({\bf x}^\prime ) ]=0$ one can
immediately infer that 
\begin{equation} 
   {\cal L}_g  \hat{\rho}_g({\bf x}) = 0 \; ,
\end{equation} 
i.e., there will be no decoherence of the 
number density at a point ${\bf x}$. This
is a reasonable result since spontaneous emission should not
change the population density if the atomic center-of-mass motion
is neglected.

To study the phase decoherence we consider the time evolution
of the field operator $\hat{\Psi}_g({\bf x})$ itself.
Introducing the operator
\begin{equation} 
  \hat{Q}({\bf x} ):= \frac{3 \gamma_0}{4\Delta^2} \int d^3x^\prime
  J(r,\theta) \Omega({\bf x}^\prime) \Omega^*({\bf x} )
   \hat{\rho}_g({\bf x}^\prime)
\end{equation} 
one can show that $ ({\cal L}_g)^n \hat{\Psi}_g({\bf x}) =
 ({\cal L}_Q)^n \hat{\Psi}_g({\bf x})$ with
\begin{equation} 
  {\cal L}_Q \hat{R} :=  \hat{Q}^\dagger({\bf x}) \hat{R}
  -  \hat{R}  \hat{Q}({\bf x})\; .
\end{equation} 
We therefore can replace ${\cal L}_g $ by ${\cal L}_Q $ for the
special case of the field operator. The solution of the equation of motion
$\partial_t \hat{R}=-{\cal L}_Q \hat{R}$ with the initial
condition $  \hat{R}(t=0) =   \hat{\Psi}_g({\bf x})$ then can be
written as
\begin{equation} 
    \hat{R}(t) =  e^{t \hat{Q}^\dagger({\bf x})}  
    \hat{\Psi}_g({\bf x})  e^{-t \hat{Q}({\bf x})}
\end{equation} 
This expression can be further reduced by writing it in the form
$ \hat{R}(t) =  \exp[-t (\hat{Q} - \hat{Q}^\dagger)]  
    \exp[t \hat{Q}] \hat{\Psi}_g({\bf x})  \exp[-t \hat{Q}]$.
The first exponential corresponds to a unitary operator. The other
two exponentials represent a non-untitary transformation of
the field operator. It is easy to prove \cite{proofsketch}
that this non-unitary transformation can be reduced to
\begin{equation} 
    e^{t \hat{Q}({\bf x})}  
    \hat{\Psi}_g({\bf x})  e^{-t \hat{Q}({\bf x})} =
    \exp \left [ - \frac{\gamma_0 t}{2} \frac{|\Omega ({\bf x})|^2 }{
    \Delta^2} \right ]  \hat{\Psi}_g({\bf x})
\label{cantrafo}\end{equation} 
This result just corresponds to the single-particle
decoherence-effect of spontaneous emission: at point ${\bf x}$
the probability for the atoms to become excited is, within
the adiabatic approximation, given by
$|\Omega ({\bf x})|^2 /\Delta^2$. The excited atoms then do
spontaneously (i.e., incoherently) decay at a rate $\gamma_0$. The
rate $\gamma_0 |\Omega ({\bf x})|^2 /\Delta^2$ therefore determines
the time scale of light-induced decoherence at point ${\bf x} $.

It is surprising that the transformation (\ref{cantrafo}) does not
describe the full decoherence of the field operator but that one
also has to take into account the unitary operator
$ U_Q := \exp[-t (\hat{Q} - \hat{Q}^\dagger)] $. If $\hat{Q}$ is Hermitean
the unitary operator becomes trivial and the total effect of decoherence
reduces to that described by Eq.~(\ref{cantrafo}). This happens
whenever the phase of the laser beam is homogeneous, for instance in
a standing-wave laser field with fixed polarization. The reason is
the cancellation between the contributions of counterpropagating
light waves. 
In a running wave, however, $\hat{Q}$ is not Hermitean and many-particle 
effects do appear.

To analyse the effect of the unitary operator $U_Q$
it is convenient to consider a fully coherent ensemble
of atoms, e.g., a Bose-Einstein condensate at temperature $T=0$.
In the mean-field description the condensate can be described 
(see, e.g., Ref.~\cite{zhang94}) by
a coherent state $|\alpha \rangle = \exp (-|\alpha|^2/2)\exp (\alpha
\hat{a}_0^\dagger)|0\rangle$, where $\hat{a}_0^\dagger = \int d^3x
\varphi_0({\bf x})  \hat{\Psi}_g^\dagger({\bf x})$ is the creation
operator for atoms in the condensate mode $\varphi_0({\bf x}) $.
The number $|\alpha|^2 $ is the number of atoms in the condensate.
Our task is then to calculate the expectation value
$\langle\alpha | \hat{R}(t)  |\alpha \rangle$ which, after some elementary
algebra, can be cast into the form
\begin{eqnarray} 
  \langle\alpha | \hat{R}(t)  |\alpha \rangle &=& \alpha \varphi_0({\bf x})
  \exp \left [ - \frac{\gamma_0 t}{2} \frac{|\Omega ({\bf x})|^2 }{
    \Delta^2} \right ] 
    \times \nonumber \\ & &
    e^{-|\alpha|^2} \langle 0 |
    e^{\alpha^* a_0} e^{\alpha \tilde{a}_0^\dagger(t)} 
    | 0 \rangle \; ,
\label{alral1} \end{eqnarray} 
where $\tilde{a}_0^\dagger (t)$ is defined as $U_Q \hat{a}_0^\dagger
U_Q^\dagger $. With techniques similar to that used to derive
Eq.~(\ref{cantrafo}) one can prove that $\tilde{a}_0^\dagger (t)$ is
in fact the creation operator for another mode,  $\tilde{a}_0^\dagger
(t) = \int d^3x \tilde{\varphi}_0({\bf x},t)  \hat{\Psi}_g^\dagger({\bf
x})$, whose mode function is given by
\begin{eqnarray}  
    \tilde{\varphi}_0({\bf x}^\prime ,t) &=& \varphi_0({\bf x}^\prime ) 
    \exp \Big [ -\frac{3}{4} i \gamma_0 t J( {\bf x} - {\bf
    x}^\prime)
    \times \label{mfunc} \\ & &\hspace{2cm}
    \frac{\Omega  ({\bf x}^\prime) \Omega^*({\bf x}) -
          \Omega^*({\bf x}^\prime) \Omega  ({\bf x})  }{i \Delta^2}
    \Big ] \; .
 \nonumber \end{eqnarray} 
Inserting this expression into Eq.~(\ref{alral1}) leads to the final
result for the light-induced decoherence of a condensate at position
$ {\bf x}$,
\begin{eqnarray} 
  \langle\alpha | \hat{R}(t)  |\alpha \rangle &=& \alpha \varphi_0({\bf x})
  \exp \left [ - \frac{\gamma_0 t}{2} \frac{|\Omega ({\bf x})|^2 }{
    \Delta^2} \right ] 
    \times\label{alral2}  \\ & &
    \exp \left [ |\alpha|^2 \left (
      \int d^3x^\prime \varphi_0^*({\bf x}^\prime) 
      \tilde{\varphi}_0({\bf x}^\prime ,t) -1 \right )
    \right ]
    \; .
\nonumber \end{eqnarray} 

\begin{figure}[t]
\epsfxsize=8cm
\epsffile{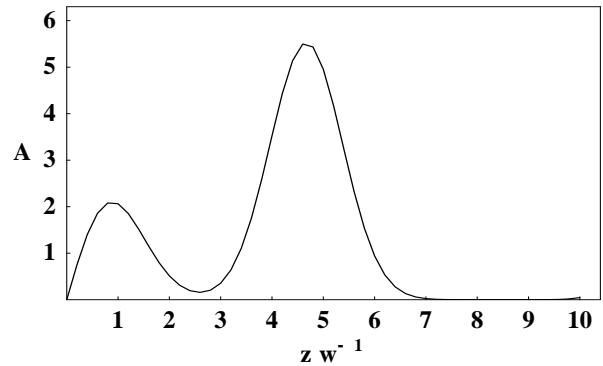}
\caption{\label{figA}The function $A(z,w,k_L)$, which is related
to a light-induced phase shift in the condensate, as a function of
the position $z$ along the z-axis. The width $w$ is $100/(2\pi)$
times the resonant wavelength, and the detuning was chosen to be
$\Delta =$ 1GHz.}
\end{figure}

Eq.~(\ref{alral2}) is the main result of this paper and describes
the total effect of light-induced decoherence on a Bose-Einstein
condensate of two-level atoms. It demonstrates that there are
non-local contributions to the atomic decoherence.
To estimate the order-of-magnitude of these non-local terms
we assume for simplicity a Gaussian condensate wave function 
of width $w$, i.e., $\varphi_0({\bf x}) = \exp [
    -{\bf x}^2 /(2w^2) ]/(w^{3/2}\pi^{3/4}) $. In addition we
consider a running-wave laser field of the form $\Omega({\bf x})=b\Omega_0
\exp (i k z )$. The analysis of Eq.~(\ref{alral2}) is then reduced
to a numerical evaluation of a complicated integral. To gain more
physical insight we expand the exponent appearing in 
Eq.~(\ref{mfunc}) to second order in $t$. Introducing the decoherence
rate $\gamma_D := \gamma_0 |\Omega_0|^2/(2 \Delta^2)$ we then can
write the result (\ref{alral2}) as 
\begin{eqnarray} 
  \langle\alpha | \hat{R}(t)  |\alpha \rangle &\approx& 
  \alpha \varphi_0({\bf x})
  \exp \Big [ -\gamma_D t (1+3 i N_\lambda A(z,w,k_L))
  \nonumber \\ & &\hspace{1cm}
  -(3\gamma_D t)^2 N_\lambda B(z,w,k_L) \Big ]
  \; .
\label{decpha}\end{eqnarray} 
Here $N_\lambda := |\alpha\varphi_0(0)|^2 \lambda_0^3$ is the number
of condensed atoms in a cube of dimension $\lambda_0^3$ at the center
of the condensate. $\lambda_0$ is the wavelength of resonant light.
The functions $A(z,w,k_L)$ and  $B(z,w,k_L)$ depend on the position
${\bf x} = z {\bf e}_z$ at which the decoherence is observed, the
width $w$ of the condensate, and on the wavevector $k_L$ of the
running laser wave. For $w = 100 c/\omega_0$ and a detuning
of $\Delta = 1$ GHz they are shown in Figs.~\ref{figA} and \ref{figB},
respectively. $A$ only has a strong dependence on $k_L$ for extreme
detunings above 1 THz while $B$ is rather insensitive to the detuning.

Eq.~(\ref{decpha}) demonstrates that the additional unitary operator
$U_Q$ does not lead to an additional exponential decay of the
coherences since to first order its effect is only a
position-dependent phase shift which is proportional to $A$. 
For longer times ($t \approx
1/\gamma_D$) the many-particle decoherence effect can become even larger
than the single-particle contribution. Since in the
example given above the function $B$ is roughly proportional to the
condensate density one arrives at the convincing result that
the many-particle decoherence grows with the atomic density.

In conclusion, we have shown that the decoherence of a gas of
ultracold atoms in a detuned laser beams is spatially varying and
depends on whether the phase of the laser beam is spatially
homogeneous. If this is not the case the one-particle spontaneous
emission process will be accompagnied by a significant many-particle
contribution with a non-exponential decay rate. This additional effect
corresponds to the expectation value of the unitary operator $U_Q$.


{\bf Acknowledgement:}
I thank J\"urgen Audretsch and Konstantin Krutitsky for many valuable
discussions. This project has been supported by the Deutsche
Forschungsgesellschaft (Forschergruppe Quantengase) and the Optik Zentrum
Konstanz. 


\begin{minipage}{8cm}
\begin{figure}[t]
\epsfxsize=8cm
\epsffile{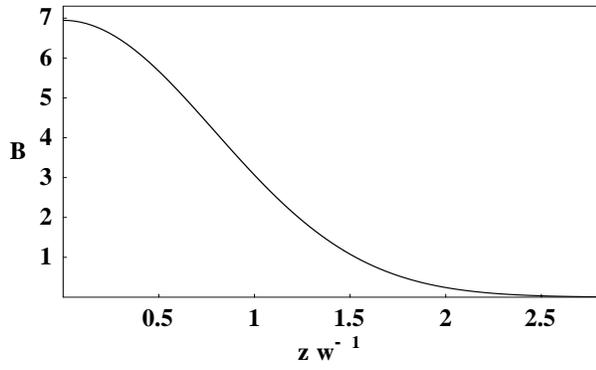}
\caption{\label{figB}The function $B(z,w,k_L)$, which is related
to the many-particle decoherence effect, for the same parameters
as in Fig.~\ref{figA}.}
\end{figure}
\end{minipage}
\end{multicols}

\begin{thebibliography}{99}
\bibitem{proofsketch} One has to show that the left-hand and the 
        right-hand side of Eq.~(\ref{cantrafo}) are the same at $t=0$
        and that they fulfill the same differential equation.
\end{thebibliography}
\end{document}